\documentclass[11pt]{article}

\usepackage[utf8]{inputenc}
\usepackage[T1]{fontenc}
\usepackage{lmodern}
\usepackage{graphicx} 
\usepackage{booktabs} 
\usepackage{array} 
\usepackage[authoryear,round]{natbib} 
\usepackage{url}
\usepackage{amsmath}
\usepackage{amssymb}
\usepackage{multirow} 
\usepackage[margin=1in]{geometry}
\usepackage{authblk}
\usepackage{float}
\usepackage{placeins}

\title{Beyond Resolution: Inference-Time Parameterized Temporal Consistency in Generative Regional Climate Downscaling}
\author[1,2]{Luca Glawion\thanks{Correspondence to: luca.glawion@kit.edu}}
\author[3]{Julius Polz}
\author[1,2]{Harald Kunstmann}
\author[1]{Benjamin Fersch}
\author[4]{Patrick Ludwig}
\author[5]{Antonio Sánchez-Benítez}
\author[1]{Christian Chwala}
\affil[1]{Karlsruhe Institute of Technology, Institute of Meteorology and Climate Research - Atmospheric Environmental Research (IMK-IFU), Garmisch-Partenkirchen, Germany}
\affil[2]{Chair of Regional Climate and Hydrology, Institute of Geography, University of Augsburg, Augsburg, Germany}
\affil[3]{Karlsruhe Institute of Technology, Institute of Meteorology and Climate Research - Atmospheric Trace Gases and Remote Sensing, Karlsruhe, Germany}
\affil[4]{Karlsruhe Institute of Technology, Institute of Meteorology and Climate Research - Troposphere Research (IMKTRO), Karlsruhe, Germany}
\affil[5]{Alfred Wegener Institute Helmholtz-Center for Polar and Marine Research, Bremerhaven, Germany }

\begin{document}
\maketitle
\vspace{-1.5cm}

\section*{Abstract}

GCM-to-RCM downscaling emulation is dominated by generative diffusion models, but most existing work evaluates skill only per timestep, treating each field as an independent snapshot. 
This leaves open whether generated sequences are temporally coherent or merely independently plausible, a gap that matters increasingly for sub-daily resolutions since downstream impacts depend on accumulation and persistence, not just on isolated states. 
This study asks whether autoregressive conditioning can deliver coherent rollouts without error accumulation and ensemble-spread collapse, and whether a single model can expose the resulting coherence-diversity trade-off as a controllable inference choice rather than a fixed design commitment.
We introduce a noise-conditioned teacher-forcing training scheme conditioning a diffusion U-Net downscaling model on a noised previous high-resolution state, so the trained model interpolates continuously between autoregressive (AR) and per-timestep independent (PTI) sampling via one noise parameter.
We apply the scheme to an hourly, European-domain historical simulation downscaling AWI-CM1 (1°) with ICON (0.11°, regridded to 0.25°) for precipitation and 2-meter temperature. The noise parameter traces a smooth interpolation between ICON's autocorrelation and PTI generation.
AR rollouts of up to six months show no progressive degradation, with ensemble spread contracting only slightly within the first 24 hours. In contrast, the same architecture trained on noise-free previous states diverges. 
For precipitation PTI underestimates maxima by up to 2x in daily aggregations, while AR tracks ICON's maxima closely. For temperature, differences between modes are negligible. 
These results show that in diffusion downscaling temporal consistency can be achieved architecture-agnostically and offered as a controllable, not fixed, inference choice.

\section{Introduction}

%% 	Climate extremes and the need for high-resolution climate information
Hydroclimatic extremes at both ends of the spectrum, floods and landslides on the one hand and water scarcity and drought on the other, have severe impacts on human societies and are projected to intensify in a warmer climate \citep{ipcc_climate_2023, ifrc_world_2021, athanase_how_2024, klimiuk_european_2025}.
These extremes are scale-dependent: sub-hourly convective bursts that cause flash floods and urban drainage failures, and multi-day persistent rain that causes fluvial floods and soil saturation, arise from fundamentally different mechanisms and act on vastly different spatial and temporal scales \citep{breugem_meteorological_2020, yang_flash_2016}.
Global Climate Models (GCMs), the primary tool for projecting climate change under warming, operate at $\approx 10$--$150$~km resolution and often do not resolve the catchment- and urban-scale rainfall processes at which impacts materialize. 
Unresolved orography, coastlines, and convective processes are known to be a primary bias source in both mean climate and extremes \citep{dawson_simulating_2015, stevens_what_2013, davini_climate_2017}.
\\\\
%%	Dynamical downscaling and its computational cost
Dynamical downscaling with regional climate models (RCMs) narrows this scale gap by integrating a physics-based limited-area model under GCM boundary conditions. 
At convection-permitting resolution this measurably improves the representation of sub-daily precipitation extremes \citep{giorgi_thirty_2019, prein_review_2015, hundhausen_sub-hourly_2025}.
This improved fidelity comes at high computational costs, limiting the exploration of different physics parametrization, warming scenarios and ensemble sizes needed to sample internal climate variability.
Addressing these limitations motivates the development of cheap, data-driven emulators that learn the GCM$\rightarrow$RCM mapping, as currently benchmarked by initiatives such as CORDEX-ML-Bench \citep{rampal_cordex-ml-bench_2026}.
\\\\
%% 	Generative AI for climate downscaling
The development of such emulators is, however, particularly challenging for precipitation, which is characterized by an intermittent, zero-inflated and heavy-tailed distribution.
Generative models have proven to be particularly effective in addressing this challenge: unlike deterministic regressors, which tend to underestimate extremes, they sample from the conditional distribution of the high-resolution fields and thereby represent uncertainty and tail behavior at a fraction of the computational cost of an RCM \citep{leinonen_stochastic_2021, harris_generative_2022, glawion_global_2025}.
With their stable, non-adversarial training and better representation of extremes, diffusion models have emerged as a leading approach, outperforming GANs and deterministic baselines in recent benchmarking \citep{rampal_cordex-ml-bench_2026}. 
This has fueled a growing interest in generative-ensemble approaches, including residual/corrective diffusion \citep{mardani_residual_2025} and diffusion and flow-matching emulators \citep{wetherell_flow_2026, addison_machine_2026, aich_conditional_2026, perkins_hiro-ace_2026}.
\\
%% 	The missing temporal dimension in generative downscaling
However, these recent efforts to advance generative AI downscaling have mostly assessed skill \textit{per timestep} only, treating each generated field as an independent snapshot.
Yet atmospheric processes and especially precipitation-relevant phenomena like fronts, mesoscale convective systems, or persistent stratiform rain, are spatiotemporal objects whose impact is governed by duration and accumulation, rather than single-timestep values. 
Hydrological and soil-moisture models integrate forcing over time and are sensitive to this structure regardless of per-timestep accuracy. 
\\\\
%% PTI versus autoregressive generation: the coherence–diversity trade-off
Generating each timestep independently of its predecessor, conditioned only on the concurrent GCM state (referred to as per-timestep independent, \textbf{PTI}, generation hereafter), is trivial to parallelize, but offers no guarantee of coherent trajectories: generated sequences may be merely independently plausible snapshots rather than physically consistent in time. 
The standard route to temporal consistency in sequence modeling is autoregressive (\textbf{AR}) conditioning, where each generated state feeds back as input to generate the next \citep{price_probabilistic_2025, lam_learning_2023, brandstetter_message_2022}. 
This closed loop exposes the model to \textit{exposure bias}: training relies on ground-truth previous states, whereas inference can only rely on the model's own, imperfect predictions, so that small deviations may compound over the rollout \citep{bengio_scheduled_2015, ranzato_sequence_2016}. 
In practice, this can lead to error accumulation and progressive degradation of generated trajectories over long rollouts, potentially manifesting as a loss of sharpness and small-scale variability \citep{bonavita_limitations_2024, couairon_archesweathergen_2026}.
% TODO: add citation for loss of sharpness / spread collapse
PTI generation sidesteps this feedback-related risk entirely, but forgoes the coherence that AR conditioning is designed to provide, resulting in a \textit{coherence--diversity trade-off} between temporally coherent trajectories and preserved ensemble spread.
\\\\
%% 	Existing approaches in generative downscaling
Existing work has addressed temporal consistency in generative downscaling in three ways. The first enforces it through the training objective: EnScale \citep{schillinger_enscale_2026} is, to our knowledge, one of the few generative RCM emulation frameworks to explicitly target temporally consistent multivariate output, using a proper scoring rule. The second imposes it at inference, e.g.\ through discriminator-based guidance of a pretrained, temporally unconditioned diffusion model \citep{hess_generating_2025}. The third generates multiple timesteps jointly \citep{glawion_global_2025, srivastava_precipitation_2024, bassetti_diffesm_2024, schmidt_generative_2025}, which is memory-intensive for long sequences at high resolution and requires stitching across generated blocks.
\\\\
%% 	Noise-conditioned autoregressive generation in video diffusion
Outside climate emulation, related approaches have been explored to achieve temporally consistent predictions in generative sequence models.
In \citet{ho_imagen_2022}, the outputs of preceding cascade stages are noise-augmented before being passed to subsequent super-resolution models, with the noise level explicitly provided to the network. \citet{blattmann_stable_2023} similarly corrupt a conditioning image at randomly sampled noise levels, but generate video frames jointly rather than autoregressively. More directly related to autoregressive generation, \citet{chen_diffusion_2024} train a causal sequence model on independently noise-corrupted frames, explicitly providing their noise levels to the network. At inference, slightly noisy previous frames can be used to stabilize long autoregressive rollouts. Related input perturbations have also been explored to improve the robustness of autoregressive idealized physics simulators \citep{sanchez-gonzalez_learning_2020}.
\\\\
However, the setting of these approaches differs fundamentally from autoregressive GCM$\rightarrow$RCM emulation. In video generation, forecasting, and PDE surrogates, the preceding state is observed at initialization and provides the dominant predictive signal. In RCM emulation, by contrast, the concurrent GCM state is the primary predictor, while the preceding high-resolution state carries only secondary, short-range persistence information. Moreover, no observed high-resolution initial state is available at rollout start. To our knowledge, noise-conditioned AR generation has not yet been investigated for RCM emulation. It therefore remains open whether this strategy can produce stable rollouts in this setting, and whether the conditioning noise level can serve as an inference-time control of the coherence--diversity trade-off.
\\\\
%% 	Research objectives, proposed approach, and research questions
We address this gap by adopting noise-augmentated AR conditioning for GCM$\rightarrow$RCM emulation, using a single trained diffusion model with a parameterized transition between AR and PTI generation at inference time. 
This is motivated by the complementary benefits of each regime: coherence on the one hand, and parallel sampling and generating initial frames without prior high-resolution states on the other. 
During training, the model is conditioned on a previous high-resolution state corrupted at a randomly sampled noise level that is not provided to the network. The model therefore learns to weigh the previous state against the concurrent GCM state by its information content.
At inference, a single noise level $\tau$ applied to the previous prediction interpolates continuously between the two regimes. 
Rollouts are initialized from the model's own PTI sample, so no observed high-resolution initial state or spin-up is required. 
Because the scheme only modifies the conditioning input and training procedure, it can be readily integrated into other conditional diffusion frameworks for sequential data generation. 
As a reference, we additionally train a conventional teacher-forcing baseline on uncorrupted previous states.
We demonstrate the approach by emulating hourly, European-domain simulations in which the ICON-CLM regional climate model ($\approx 12$~km) downscales an AWI-CM1 global simulation ($\approx 1^\circ$) spectrally nudged toward ERA5. Both datasets are regridded to a common $0.25^\circ$ grid, and we focus on precipitation and 2-meter temperature, evaluating the emulation on a held-out six-month period (July--December 2021).
While this setup is not convection-permitting, the training scheme makes no assumption about spatial or temporal resolution.
\\\\
Our study is guided by three questions. 
First (\textit{feasibility and control}), can AR conditioning produce temporally coherent rollouts when the concurrent GCM state, rather than the previous state, is the primary predictor and no observed high-resolution initial state is available, and can a single trained model provide continuous inference-time control over the coherence--diversity trade-off, interpolating between AR and PTI generation without retraining? 
Second (\textit{rollout robustness}), does coherence come at the characteristic cost of exposure bias, i.e.\ error accumulation, smoothing, or reduced or miscalibrated ensemble spread? 
Third (\textit{downstream relevance}), how does the presence or absence of temporal consistency affect relevant statistics from hourly extremes to multi-day accumulations?

\begin{figure}
    \centering
    \includegraphics[width=1\linewidth]{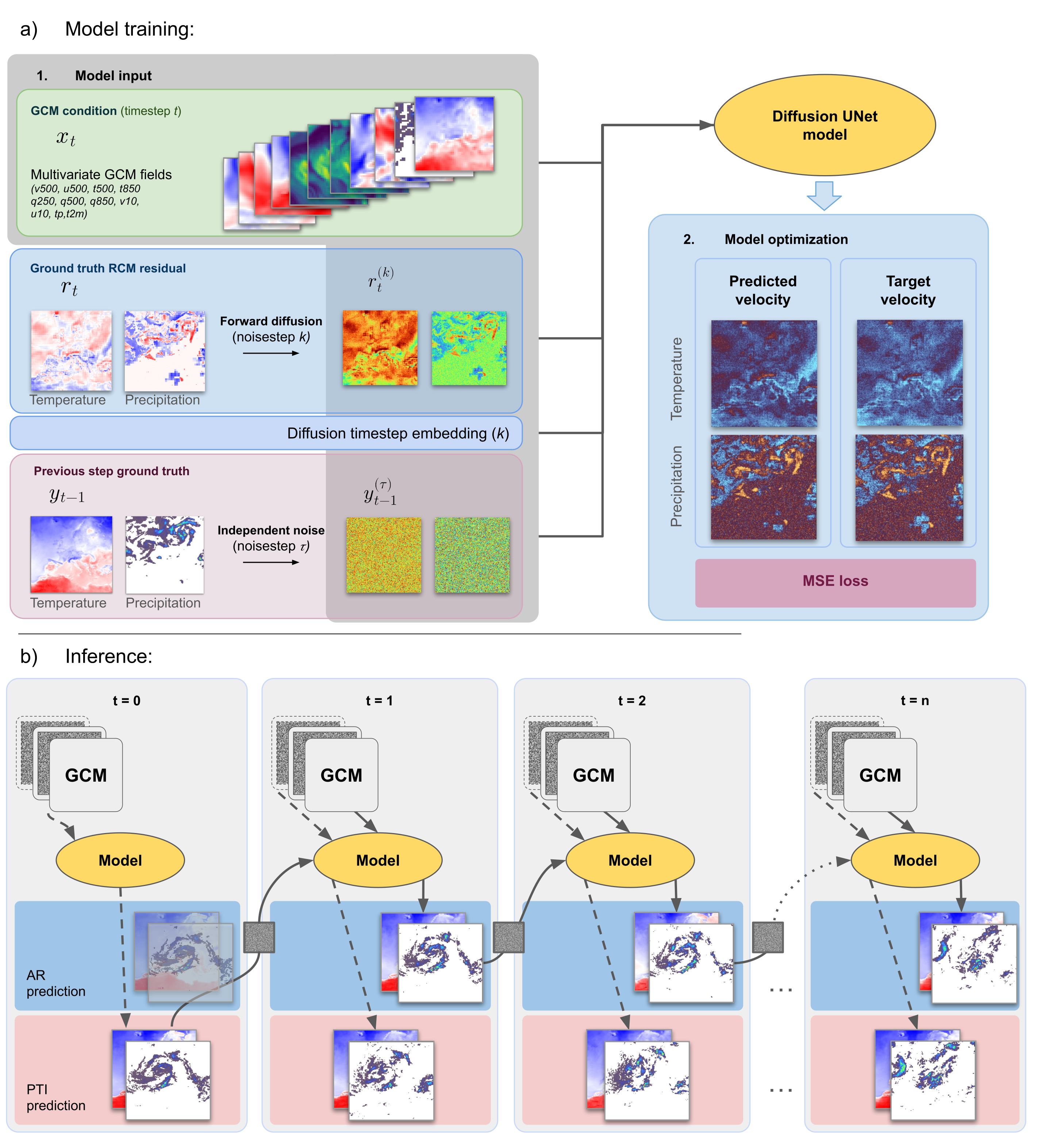}
    \caption{Overview of the proposed temporal-noise-conditioned diffusion framework. (a) Training: The conditional diffusion model learns to emulate high-resolution RCM fields from coarse GCM input. The previous-hour RCM state is additionally provided as temporal condition input, corrupted with randomly sampled noise. (b) Inference: 
The conditioning noise level is controlled by the parameter \(\tau\). At its maximum value (\(\tau=999\)), the temporal condition consists of pure noise, resulting in per-timestep independent (PTI) samples (dashed lines). This regime also serves to initialize the autoregressive rollout. At a lower noise level (e.g. \(\tau=50\)), the previous timestep's prediction is perturbed with noise and used as the temporal conditioning input. This provides informative temporal context, resulting in temporally consistent autoregressive (AR) sampling (solid lines).
}
    \label{fig:overview}
\end{figure}

\section{Temporal-Noise-Conditioned Diffusion Emulation: \\ Methodology \& Data}

\subsection{Overview}
We develop a conditional diffusion model for statistical emulation of regional climate model (RCM) output from global climate model (GCM) fields, using a novel noise-conditioned training strategy to enable continuous control over temporal conditioning. Figure \ref{fig:overview} gives an overview of the model input, the training strategy, and how the same trained model is applied in autoregressive or per-timestep independent mode during inference.
The network predicts the residual between the RCM fields of 2\,m temperature (t2m) and total precipitation (tp) and the corresponding coarse GCM input. An additional conditioning channel carries the previous hour's RCM state, corrupted by forward-diffusion noise of randomly sampled strength during training. During inference, the noise level $\tau$ of this conditioning input controls the degree of temporal coupling, enabling a continuous interpolation between AR and PTI generation with the same trained weights. The input and target data comprise hourly output (2018--2024) from an AWI-CM1 storyline simulation ($\approx 1^\circ$) and its ICON-CLM downscaling over Europe ($\approx 12$\,km), both conservatively remapped to a $0.25^\circ$ grid. This resolution provides a computationally tractable setup for investigating the proposed temporal conditioning scheme. The year 2021 is held out for validation and testing.

\subsection{Temporal-Noise-Conditioned Diffusion Model}
\label{sec:temporal-conditioning}

Let \(x_t\) and \(y_t\) denote standardized GCM input and RCM target at hour $t$, respectively, with both fields represented on the same \(176\times176\) grid (Section~\ref{sec:data-preprocessing}). We define the residual fine-scale target $r_t = y_t - P\,x_t$ as the difference between the RCM and GCM 2-meter temperature (t2m) and total precipitation (tp) fields, where $P$ selects the t2m and tp channels of $x_t$. The diffusion model predicts \(r_t\), from which the standardized RCM fields are reconstructed as \(y_t=r_t+P\,x_t\). Finally, the physical fields are obtained by applying the inverse RCM standardization.
\\\\
The conditional diffusion model generates $r_t$ conditioned on the coarse state $x_t$ and on a noise-corrupted version of the previous-hour RCM state $y_{t-1}$ (the standardized RCM field rather than the residual). The target residual and the previous state are corrupted by the same forward-diffusion process, but at two independent noise steps $k$ and $\tau$,
\begin{align}
r_t^{(k)} &= \sqrt{\bar\alpha_k}\,r_t + \sqrt{1-\bar\alpha_k}\,\epsilon, & \epsilon &\sim \mathcal{N}(0,\mathbf{I}), \label{eq:forward}\\
y_{t-1}^{(\tau)} &= \sqrt{\bar\alpha_\tau}\,y_{t-1} + \sqrt{1-\bar\alpha_\tau}\,\epsilon', & \epsilon' &\sim \mathcal{N}(0,\mathbf{I}), \label{eq:tc}
\end{align}
where $k,\tau\in\{0,\dots,T-1\}$, $T=1000$, $\bar\alpha$ is the cumulative signal coefficient of a cosine noise schedule \citep{ho_denoising_2020, nichol_improved_2021}, and $\epsilon$ and $\epsilon'$ are independent. The diffusion step $k$ indexes the generative process of the target, i.e.\ $q(r_t^{(k)}\mid r_t)=\mathcal{N}(\sqrt{\bar\alpha_k}\,r_t,\,(1-\bar\alpha_k)\,\mathbf{I})$. The conditioning-noise step $\tau$ instead controls how much information about the previous state is retained: for small $\tau$, it is close to $y_{t-1}$, whereas for $\tau$ close to $T-1$ it is almost pure noise. The network $v_\theta$ is trained to predict the velocity $v$ \citep{salimans_progressive_2022} and is optimized with
\begin{equation}
v=\sqrt{\bar\alpha_k}\,\epsilon-\sqrt{1-\bar\alpha_k}\,r_t, \qquad
\mathcal{L}(\theta)=\mathbb{E}_{r_t,\epsilon,\epsilon',k,\tau}\Bigl[\bigl\lVert v_\theta\bigl(r_t^{(k)},k,x_t,y_{t-1}^{(\tau)}\bigr)-v\bigr\rVert^2\Bigr],
\label{eq:loss}
\end{equation}
the mean squared error over both target channels and all grid points. Only $k$ is provided to the network. The conditioning-noise step $\tau$ is not, so the network must learn to adapt its use of the conditioning state to its information content.
\\\\
The network $v_\theta$ is a two-dimensional U-Net implemented with Diffusers \citep{von_platen_diffusers_2022, ronneberger_u-net_2015} with four resolution levels (64, 128, 128 and 256 channels), two residual blocks per level and self-attention at resolutions of $44\times44$ and $22\times22$. The diffusion step $k$ is provided through a sinusoidal embedding. The network has $17.8$\,million parameters. Its input is the channel-wise concatenation $[\,r_t^{(k)}\,|\,x_t\,|\,y_{t-1}^{(\tau)}\,]$, and its output is the velocity of both target channels.
\\\\
Training and inference share equations~\eqref{eq:forward}--\eqref{eq:loss} and the network weights. The two phases differ only in three respects: (i) the previous state $y_{t-1}$ is the observed RCM state during training and the model's own previous prediction $\hat y_{t-1}$ during inference; (ii) the conditioning-noise step $\tau$ is drawn at random for every training sample but fixed to a user-chosen value $\tau=s$ during inference; and (iii) during training $r_t^{(k)}$ is obtained by noising the true residual with equation~\eqref{eq:forward}, whereas during inference it is initialized with pure noise and iteratively denoised to the prediction $\hat r_t$. These differences are described in detail in the following two sections.

\subsection{Training: Teacher Forcing with Randomized Conditioning Noise}

During training, $y_{t-1}$ is the true RCM state of the previous hour, which is analogous to teacher forcing. However, instead of supplying the previous state unchanged, we corrupt it with equation~\eqref{eq:tc} at a conditioning-noise step $\tau\sim\mathcal{U}\{0,\dots,T-1\}$ sampled uniformly over the full diffusion schedule, exposing the network to previous states ranging from nearly clean to almost pure noise. The target is noised with equation~\eqref{eq:forward} at a uniformly drawn diffusion step $k$. The quantities $k$, $\tau$, $\epsilon$ and $\epsilon'$ are drawn independently of each other and anew for every sample. Because the conditioning always uses observed states, training requires neither model-generated previous states nor iterative autoregressive rollouts.
\\\\
We trained for 100 epochs with AdamW (weight decay 0.01) \citep{loshchilov_decoupled_2019} at a learning rate of $10^{-5}$ (linear warm-up from $10^{-6}$ over 8 epochs, constant for 35 epochs, then cosine cycles between $10^{-5}$ and $10^{-6}$ with a period of 50 epochs), gradient-norm clipping at 1.0 and a batch size of 16 per GPU on four GPUs (A100).  An exponential moving average (EMA) of the model weights was maintained with a decay of 0.95, updated after every optimizer step. The EMA weights were used for validation and inference. We selected the checkpoint with the lowest validation loss, i.e.\ the mean squared error of the velocity prediction on the validation set. The model was implemented in PyTorch with the diffusers library \citep{paszke_pytorch_2019, von_platen_diffusers_2022}.

\subsection{Inference: Autoregressive Rollout with Adjustable Conditioning-Noise Level}

Ensemble members are generated independently, in chronological order. For each member, the model's own previous prediction $\hat y_{t-1} = \hat r_{t-1} + P\,x_{t-1}$ takes the place of $y_{t-1}$ in equation~\eqref{eq:tc} \citep{rasul_autoregressive_2021}, and the conditioning-noise step is fixed to $\tau=s$, with a new draw of $\epsilon'$ at every hour. The residual $\hat r_t$ is generated by reverse diffusion from pure noise given $x_t$ and $\tilde y_{t-1}$, and $\hat y_t = \hat r_t + P\,x_t$ is passed on to the next hour. In the first hour, which has no previous prediction, $\tilde y_{t-1}$ is drawn from $\mathcal{N}(0,\mathbf{I})$, practically equivalent to $\tau=999$.
\\\\
The conditioning-noise level $\tau$ is fixed throughout each rollout and controls the information retained from the previous prediction. 
We refer to sampling with a strongly informative previous-state condition as autoregressive sampling and to sampling with the previous-state information largely suppressed as weakly coupled sampling. In the latter regime, temporal dependence can still arise through the evolving GCM forcing \(x_t\).
We denote the output generated with conditioning-noise level $\tau$ by $\mathrm{AR}_\tau$. We focus on two settings: $\mathrm{AR}:=\mathrm{AR}_{50}$ ($\tau=50$), whose condition remains close to $\hat y_{t-1}$ and anchors the model to its own recent past, which yields temporally coherent trajectories, and the per-timestep independent setting $\tau=999$, hereafter PTI, in which information from the previous prediction is suppressed. Both settings use the same trained weights and differ only in $\tau$. We additionally evaluate intermediate values of $\tau$ (e.g.\ $\mathrm{AR}_{200}$) to investigate the continuous interpolation between strongly autoregressive and weakly coupled sampling.
\\\\
Each field is generated with 30 steps of the deterministic DDIM sampler \citep{song_denoising_2021} instead of the 1,000 training steps, using the $v$ parameterization and clipping the predicted clean residual to $[-2,2]$ at every step; the reverse process injects no noise. The initial noise of the reverse process and $\epsilon'$ are drawn anew for every hour and member, and each member feeds back its own prediction and thus follows its own trajectory. 

\subsection{Conventional Teacher-Forcing Baseline}
\label{sec:tf-baseline}

As an additional baseline, we train the conditional diffusion emulator with conventional teacher forcing \citep{bengio_scheduled_2015}. Teacher forcing conditions a sequence model on the ground-truth previous state during training, whereas at inference the model's own prediction is fed back. This train--inference mismatch (exposure bias) can let errors accumulate quickly once the model drifts into states it never encountered during training, yet the strategy remains common in autoregressive emulators \citep{price_probabilistic_2025, huang_generative_2026}. The baseline uses the same emulator setup as described in Section~\ref{sec:temporal-conditioning}, except that the previous RCM state enters the network without noise, i.e.\ $y_{t-1}$ takes the place of $y_{t-1}^{(\tau)}$ in the network input during training. At inference, the previous model prediction $\hat y_{t-1}$ is likewise fed back without perturbation. Since this model cannot generate the first state itself, the rollout is initialized with the RCM field of the first hour of the test period. We denote outputs generated with the baseline model by $\mathrm{TF}$.

\subsection{Data}

The GCM data driving our emulator come from a coupled-model storyline simulation performed with the CMIP6 model AWI-CM-1-1-MR \citep{semmler_simulations_2020}, in which the atmospheric model ECHAM6.3.04p1 (T127L95 spectral resolution, $\approx$100~km, 95 vertical levels) is coupled to the FESOM v1.4 ocean/sea-ice model ($8$--$80$~km mesh, refined in dynamically active regions such as the Gulf Stream). Large-scale free-tropospheric (700~hPa--100~hPa) divergence and vorticity are spectrally nudged, toward ERA5 reanalysis \citep{hersbach_era5_2020}, retaining only zonal wavenumbers up to 20, while thermodynamics, the atmospheric boundary layer and small-scale dynamics evolve freely, letting the model's own physics govern near-surface variables such as temperature and precipitation. See more details about the configuration in \citet{klimiuk_european_2025} and \citet{athanase_projected_2024}. For this study, we use a \textit{present-day} simulation (2018--2024, hourly output) as GCM-side conditioning input on AWI-CM1's native global grid ($\approx 1^\circ$).
\\\\
Our regional climate (RCM) target comes from dynamically downscaling the AWI-CM1 storylines with ICON-CLM \citep{klimiuk_european_2025}, the climate-mode configuration \citep{pham_icon_2021} of the ICOsahedral Nonhydrostatic (ICON) model \citep{zangl_icon_2015}, run on the EURO-CORDEX domain (rotated lat--lon grid) at R12B5 ($\approx$12~km, $0.11^\circ$) and hourly resolution. Boundary forcing comes from the nudged AWI-CM1 fields, with additional grid-point nudging of horizontal velocity at the upper boundary (starting at 10,500,m) to keep the regional circulation close to its driving global simulation. Nudging coefficients for thermodynamic variables (virtual potential temperature, density, specific humidity) are set to zero so temperature, moisture, and precipitation develop freely from the regional model's own physics. For this study the full-period (2018--2024) RCM downscaled fields matching the GCM simulation described above were available.

\subsection{Preprocessing, Normalization and Data Selection}
\label{sec:data-preprocessing}

GCM and RCM fields were conservatively remapped onto the same $0.25^\circ$ rotated lat--lon grid of $188\times183$ points preserving area-integrated quantities. The model operates on a $176\times176$ crop of this grid. The GCM fields were then averaged over blocks of $4\times4$ grid points ($\approx 1^\circ$, the native GCM resolution) and replicated onto the $0.25^\circ$ grid by nearest-neighbor upsampling, so that the coarse input is piecewise constant on $4\times4$ blocks and shares the grid of the target. GCM pressure-level fields, which are available at 3-hourly intervals, were interpolated linearly in time to match the hourly resolution of the surface fields. Total precipitation is defined as the sum of large-scale and convective precipitation (mm\,h$^{-1}$).
\\\\
The input $x_t$ comprises 11 channels of the regridded GCM: t2m, tp, the 10\,m wind components (u10, v10), specific humidity q (850, 500 and 250\,hPa), temperature t (850 and 500\,hPa) and the wind components u and v (500\,hPa), reflecting the commonly used combination of near-surface meteorological and large-scale atmospheric predictors in climate downscaling \citep{rampal_cordex-ml-bench_2026}. The targets are t2m and tp of the RCM. Non-negative variables (precipitation and humidity) were clipped at zero. All variables were standardized to zero mean and unit variance (z score), except precipitation, which was log-transformed, $\log(p+\varepsilon)-\log\varepsilon$ with $\varepsilon=0.01$\,mm\,h$^{-1}$, and then min--max scaled to $[-1,1]$. Normalization statistics were computed separately for the GCM and the RCM. 
The residual $r_t = y_t - P\,x_t$ was formed after normalization, as the difference between the normalized RCM target $y_t$ and the corresponding normalized GCM field $P\,x_t$ for both t2m and tp; the temporal condition uses the normalized RCM state $y_{t-1}$. Predictions are mapped back by adding $P\,x_t$, inverting the normalization (inverse min--max scaling and inverse log transform for precipitation) and clamping precipitation at zero, with values below $0.01$\,mm\,h$^{-1}$ set to zero.
\\\\
We split the 2018--2024 period into non-overlapping training (2018--2020 and 2022--2024; 51,864 hourly samples), validation (1--30 March 2021) and test (1 July--31 December 2021) sets.
For the test period, predictions were generated using both PTI and AR approaches. The AR predictions were initialized on 1 July 2021. For the rollout evaluation, additional AR initializations at weekly intervals were considered. In all AR and PTI experiments, ensembles of 12 members were generated. The TF baseline was rolled out over the full test period with a single member.
\\\\
To evaluate temporal consistency, the autocorrelation of 2\,m temperature is analyzed using anomalies relative to a time-of-day climatology, since the seasonal and diurnal cycles would otherwise mask differences in temporal consistency. The temperature anomaly is defined for each RCM grid pixel: for each calendar month, we compute the mean value at each hour of day over all training-set years and subtract this per-pixel, per-month, per-hour climatology from the raw temperature series.

\subsection{Evaluation Metrics}

All metrics are computed in physical units on hourly fields, unless temporal aggregation is indicated. Deterministic metrics use the first ensemble member, whereas probabilistic scores and rollout diagnostics use all \(N=12\) members \(\hat y_t^{(m)}\), which are compared with the RCM target \(y_t\).
\\
Rollout stability and ensemble calibration are assessed with the root mean squared error \(\mathrm{RMSE}_t\) of the ensemble mean \(\overline{\hat y}_{t,i}\) over all grid cells \(i\in\Omega\), together with the ensemble spread and the spread--skill ratio (SSR),
\begin{equation}
\mathrm{Spread}_t=\sqrt{\frac{1}{|\Omega|}\sum_{i\in\Omega}\frac{1}{N-1}\sum_{m=1}^{N}\bigl(\hat y_{t,i}^{(m)}-\overline{\hat y}_{t,i}\bigr)^2},
\qquad
\mathrm{SSR}_t=\sqrt{\frac{N+1}{N}}\,\frac{\mathrm{Spread}_t}{\mathrm{RMSE}_t}.
\end{equation}
Because the ensemble variance is averaged over \(\Omega\) before taking the square root, spread and RMSE are directly comparable second-moment quantities, and the prefactor corrects for the finite ensemble size \citep{fortin_why_2014}. A reliable ensemble yields \(\mathrm{SSR}=1\), whereas \(\mathrm{SSR}<1\) indicates underdispersion.
\\\\
Temporal consistency is quantified by the lagged autocorrelation function (ACF), computed at each grid cell and then averaged over the domain. For precipitation, the ACF is computed from the hourly series directly, and for 2\,m temperature from anomalies relative to the time-of-day climatology (Section~\ref{sec:data-preprocessing}), so that the seasonal and diurnal cycles do not dominate the signal.
\\\\
Spatial variability is assessed with the radially averaged power spectral density (RAPSD) \citep{harris_multiscale_2001, sinclair_empirical_2005}, obtained by averaging the two-dimensional Fourier power spectrum of each field over annuli of integer wavenumber and then over time. GCM spectra are computed on the native \(\approx 1^\circ\) grid, i.e.\ after \(4\times4\) average pooling. As a scalar measure of spectral error, the RAPSD distance of each member to the RCM is defined as the mean absolute difference of the logarithmic spectra over all non-zero wavenumbers.
\\\\
Ensemble reliability is evaluated with rank histograms \citep{candille_evaluation_2005, hamill_interpretation_2001} of the RCM target among the sorted ensemble members, pooled over all grid cells and time steps. Ties, which are frequent for zero precipitation, are broken by drawing the rank uniformly among the tied positions. For a reliable ensemble, all \(N+1\) ranks are equally likely.

\section{Results}

This section focuses on a comparison of AR and PTI generation of precipitation and temperature fields. The conventional teacher-forcing baseline (TF) is only shown in Section~\ref{sec:accumulation}. If not further specified the underlying data is the 6 month test period from 1st July to 31st December 2021 and AR refers to strong temporal conditioning ($\tau=50$) while PTI refers to no retained temporal conditioning ($\tau=999$). Section \ref{sec:pers} shows how different values of $\tau$ affect the autocorrelation of fields.

\subsection{Case Study: Single-Event Evolution}
\label{sec:case}

\begin{figure}[!htbp]
    \centering
    \includegraphics[width=1\linewidth]{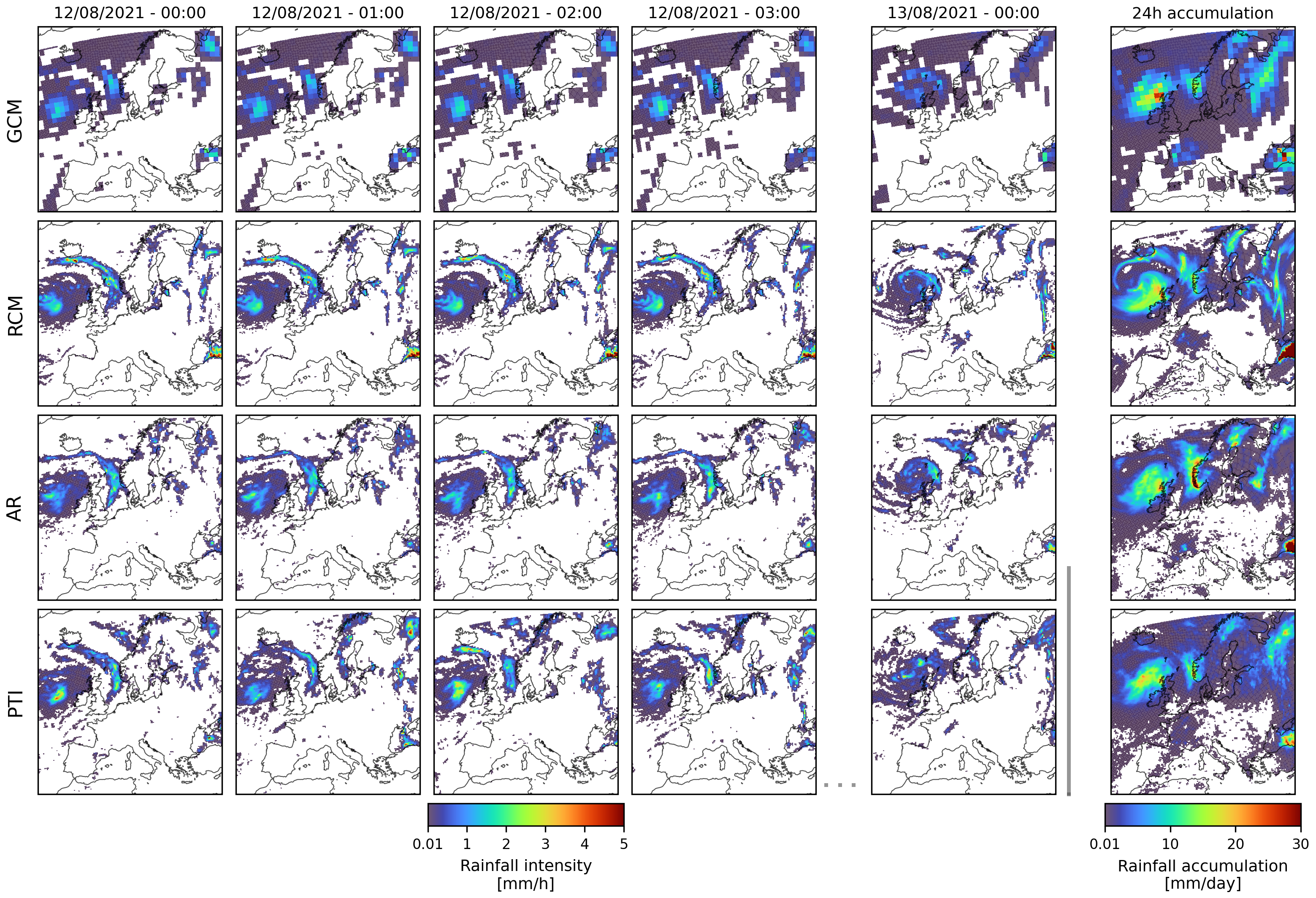}
    \caption{Hourly precipitation (mm/h) for a rain event evolving over 12--13 August 2021 (00:00--03:00 on 12 August and 00:00 on 13 August shown), with the corresponding 24-hour aggregation (mm/day, rightmost column), for the GCM (conditioning input, top row), the RCM (reference, 2nd row), the AR Emulator (3rd row) and the PTI Emulator (bottom row). }
    \label{fig:single_event_agg}
\end{figure}

\begin{figure}[!htbp]
    \centering
    \includegraphics[width=1\linewidth]{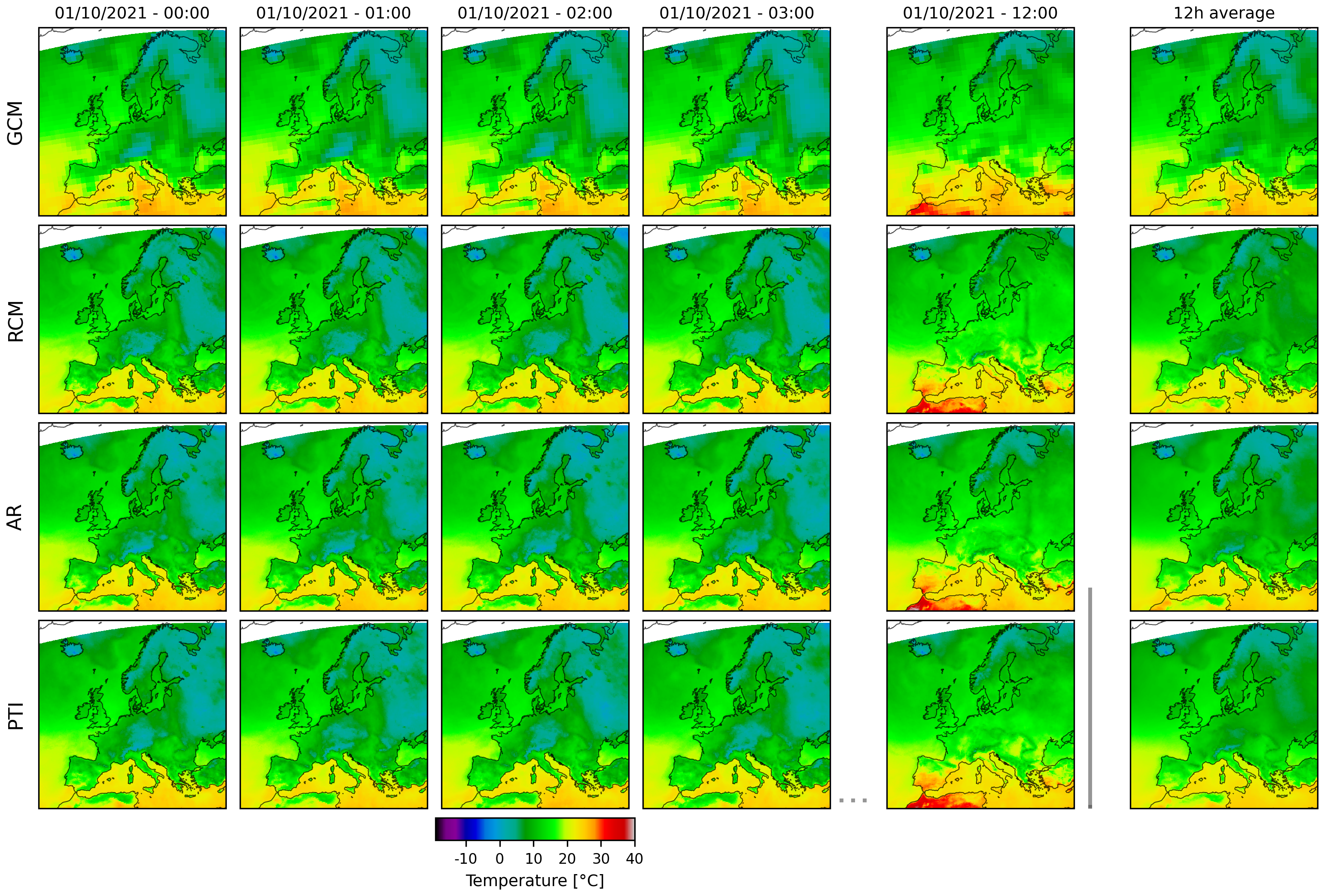}
    \caption{2 m temperature (°C) for an event evolving over 1 October 2021 (00:00--03:00 and 12:00 shown), with the corresponding 12-hour average (00:00--12:00 inclusive, rightmost column), for the GCM (conditioning input, top row), the RCM (reference, 2nd row), the AR Emulator (3rd row) and the PTI Emulator (bottom row).}
    \label{fig:single_event_t2m}
\end{figure}

Figure~\ref{fig:single_event_agg} shows hourly precipitation for a representative rain event (12--13 August 2021) alongside the corresponding 24-hour aggregation, for the GCM, RCM, AR Emulator ($\tau=50$, initialized July 1st, 00:00), and PTI Emulator ($\tau=999$). Both emulator modes reproduce precipitation fields of plausible visual quality at every individual timestep and broadly track the large-scale structure present in the GCM. However, the two modes differ markedly in how the event evolves across consecutive hours: the AR Emulator maintains coherent, slowly evolving precipitation maxima whose location and shape change smoothly from one frame to the next, closely resembling the RCM's evolution (see animation of the full event in the ancillary files). The PTI Emulator, in contrast, produces maxima that appear, disappear, and relocate between successive hourly frames, despite each individual frame looking locally realistic. This difference propagates to the 24-hour aggregation: the AR Emulator's aggregated maxima coincide in magnitude and location with the RCM, whereas the PTI Emulator's aggregation underestimates peak accumulated precipitation, as spatially inconsistent per-timestep maxima fail to stack. This effect is especially evident over the Black Sea where the AR Emulator and the RCM both show a stationary rain event that keeps relocating in the PTI Emulator frames leading to underestimation.
\\\\
Figure~\ref{fig:single_event_t2m} shows the single-event comparison for 2 m temperature, but for a different date (1 October 2021): hourly fields at 00:00--03:00 and at 12:00, alongside the corresponding 12-hour average (00:00--12:00 inclusive), for the GCM, RCM, AR Emulator, and PTI Emulator. Unlike the precipitation case, the per-timestep fields are visually more similar across the AR and PTI Emulators at every hour shown, including the pronounced midday warm anomaly over the Iberian Peninsula. This similarity persists in the 12-hour average: no equivalent of the precipitation aggregation gap between modes is visually apparent here.

\subsection{Six-Month Accumulation and Teacher-Forcing Baseline}
\label{sec:accumulation}

\begin{figure}[!htbp]
    \centering
    \includegraphics[width=1\linewidth]{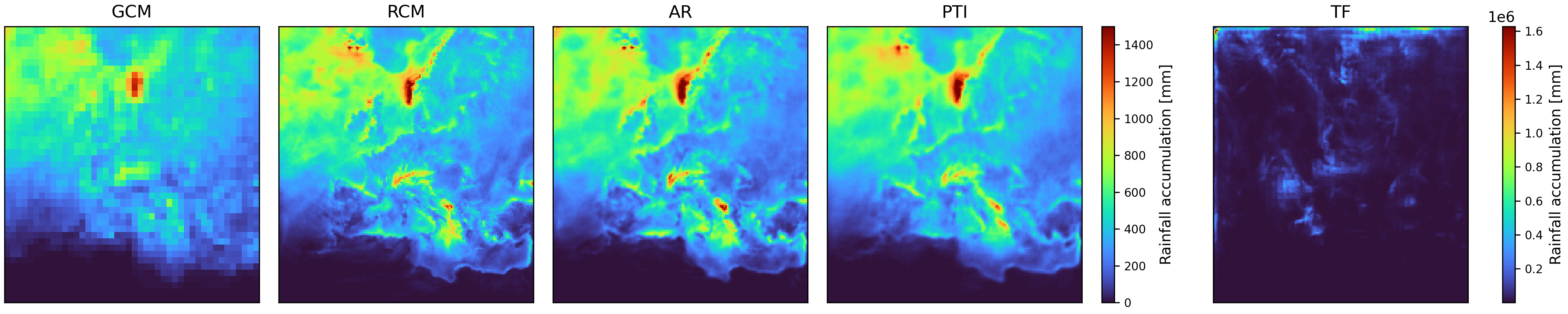}
    \caption{Precipitation accumulated over the six-month test period (1 July--31 December 2021) for the GCM, RCM, AR Emulator and PTI Emulator, together with the conventional teacher-forcing baseline (TF, Section~\ref{sec:tf-baseline}), which was trained on uncorrupted previous states and initialized with the RCM field of the first hour.}
    \label{fig:map_sum}
\end{figure}

Extending the 24-hour aggregation of the case study to the full test period, Figure~\ref{fig:map_sum} shows precipitation accumulated over six months. Both the AR and PTI Emulators reproduce the large-scale spatial distribution of European precipitation reasonably well, although the RCM retains somewhat finer-scale structures. The PTI fields are generally smoother than those produced by the AR Emulator. The AR rollout runs uninterrupted over the entire period, starting from its own PTI sample, and remains stable throughout.
\\\\
The conventional teacher-forcing baseline (TF, Section~\ref{sec:tf-baseline}) behaves fundamentally differently, even though it starts from the RCM field of the first hour rather than from a generated state. Its autoregressive rollout becomes unstable, with errors building up especially close to the domain boundaries and accumulated totals exceeding $10^6$\,mm. We therefore exclude TF from the subsequent analyses, which compare the AR and PTI Emulators only.

\subsection{Temporal Consistency}
\label{sec:pers}
\begin{figure}[!b]
    \centering
    \includegraphics[width=0.9\linewidth]{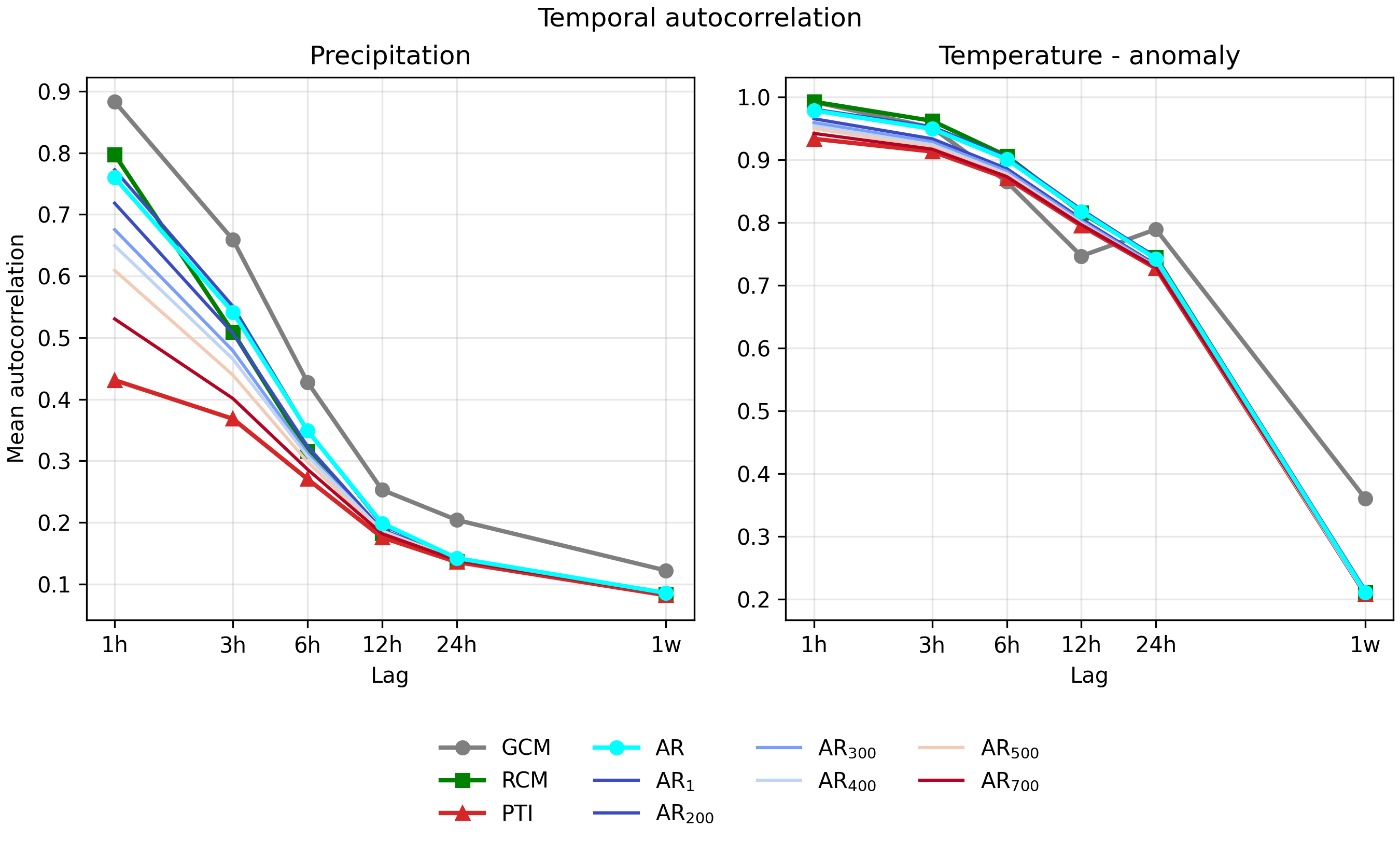}
    \caption{Lagged (1~h--1~week) autocorrelation of precipitation (left) and temperature anomaly (right), computed over the full test set and domain, for the GCM, RCM, PTI Emulator, and AR Emulator at noise levels $\tau\in\{1,50,200,300,400,500,700\}$.}
    \label{fig:acf_noise}
\end{figure}
Figure~\ref{fig:acf_noise} shows lagged autocorrelation (1~h to 1~week) of precipitation and temperature anomaly (Section~\ref{sec:data-preprocessing}), averaged across the full test period (2021-07-01  -  2021-12-31) and domain, for the GCM (interpolated), RCM, the PTI Emulator ($\tau=999$), and the AR Emulator run at seven fixed noise levels $\tau\in\{1,50,200,300,400,500,700\}$ (note that $\tau=50$ is the standard `AR Emulator` setting). For precipitation, autocorrelation decreases monotonically with $\tau$ at every lag: the lowest-$\tau$ runs exceed RCM autocorrelation at lags beyond 1~h with slightly better results for $\tau=50$ than $\tau=1$, while the PTI Emulator falls below the RCM at all lags up to 12~h, with the gap vanishing by 1 week. Intermediate $\tau$ values trace a smooth, ordered interpolation between these two extremes at every lag, with no crossovers between curves.
\\\\
Temperature anomaly autocorrelation is substantially higher than for precipitation and the spread across $\tau$ is correspondingly compressed, though the same ordering (lower $\tau$ $\to$ higher autocorrelation) holds. All curves except the GCM nearly converge by the 24~h lag, with the offset (especially at a 12~h lag) reflecting the use of the RCM climatology for anomaly calculation of the GCM not fully removing the daily cycle.
Given this early convergence and the comparatively compressed spread across $\tau$, the remaining analyses focus on precipitation, where inter-mode differences are larger and persist over longer timescales. Temperature anomaly results serve here as additional evidence that the noise-level mechanism governs temporal consistency across variables and to demonstrate the model's multivariate downscaling capability.

\subsection{Spectral Fidelity and Extremes Under Temporal Aggregation}
\begin{figure}[!htbp]
    \centering
    \includegraphics[width=1\linewidth]{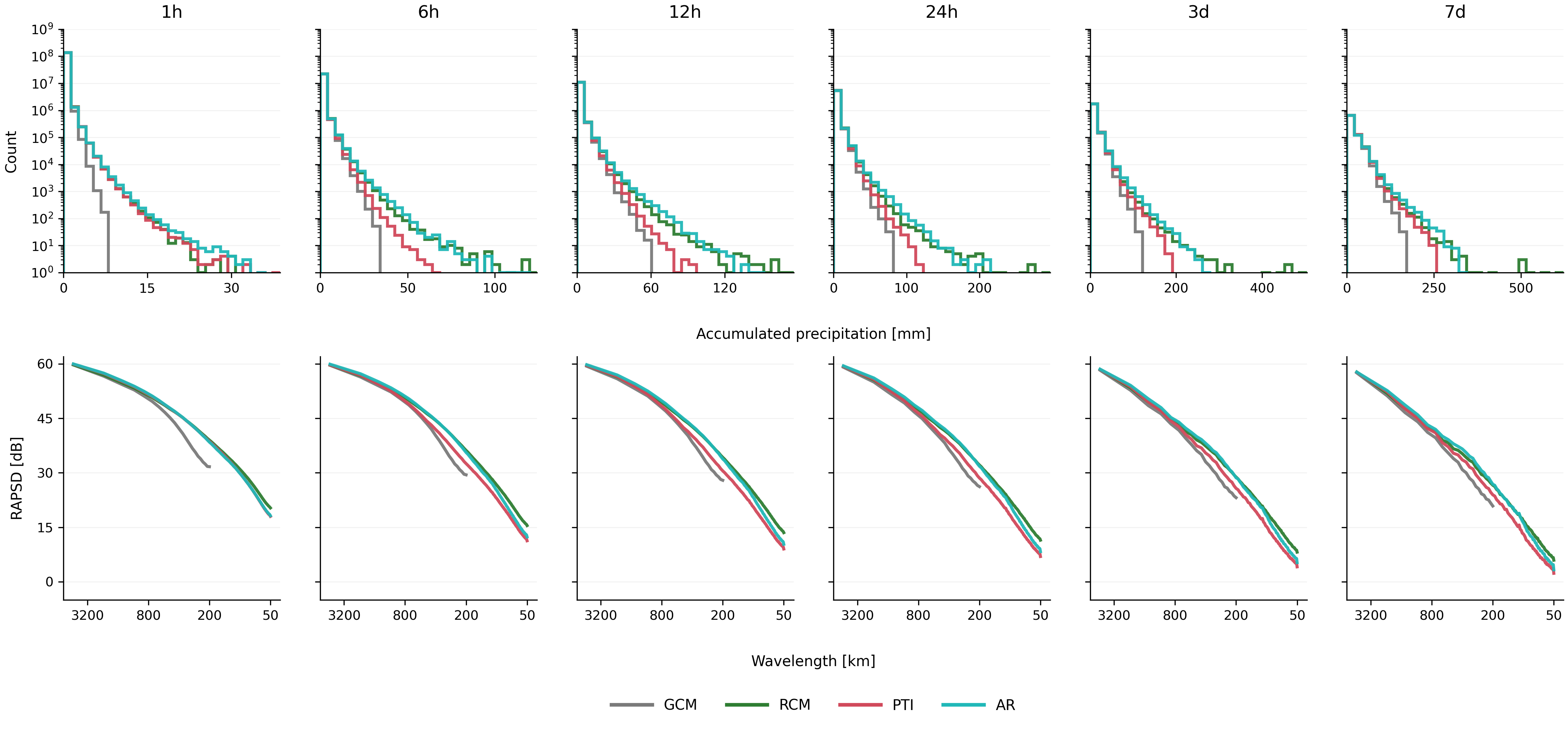}
    \caption{Precipitation histograms and power spectra at increasing temporal aggregation (1~h $\rightarrow$ multi-day) for GCM, RCM, and both Emulator modes.}
    \label{fig:hist_rapsd}
\end{figure}
\begin{figure}[!htbp]
    \centering
    \includegraphics[width=1\linewidth]{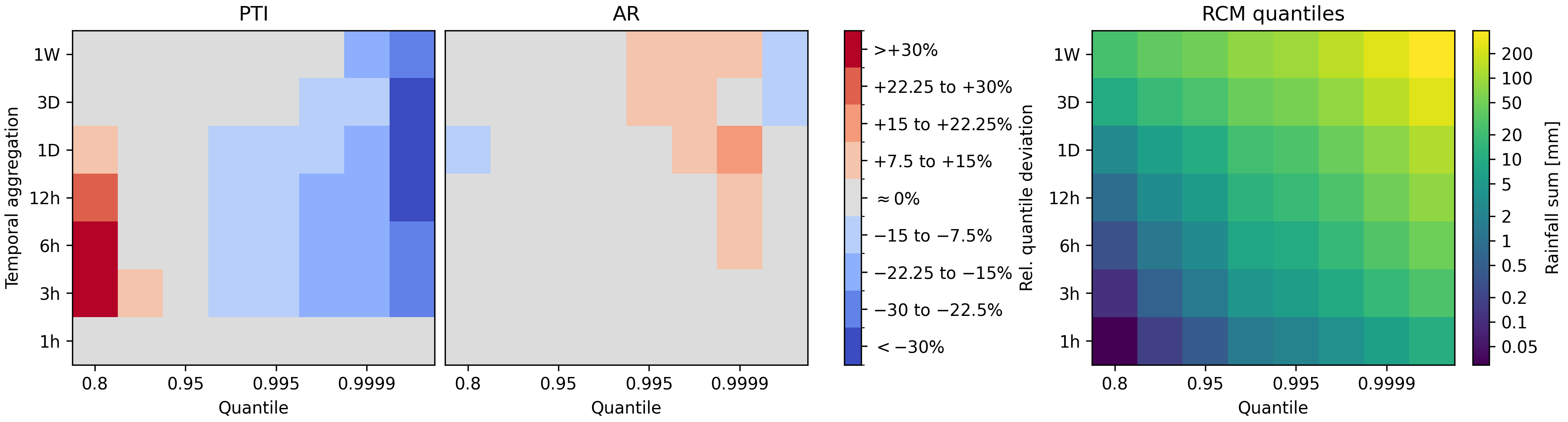}
    \caption{Relative quantile deviation from RCM (both Emulator modes) across temporal aggregations, with RCM quantile magnitudes shown for reference.}
    \label{fig:quantile_agg}
\end{figure}

At 1 h, the PTI Emulator closely reproduces the RCM precipitation histogram, while the AR Emulator slightly overestimates the heavy tail (see Figure \ref{fig:hist_rapsd}.) The power spectra nevertheless show both emulator modes exhibiting a similar fine-scale deficit relative to the RCM. Under temporal aggregation, the histogram gap between the PTI Emulator and RCM/AR Emulator widens progressively, reflecting an increasingly pronounced underestimation of precipitation extremes by the PTI Emulator. All three distributions reconverge only beyond roughly three days.
\\\\
The RAPSD gap between the PTI Emulator and AR Emulator, indistinguishable at 1~h, likewise opens only under temporal averaging. The RAPSD deficit of the AR mode is stationary under averaging while in the PTI case the opening gap becomes stationary beyond 6 hours.
\\\\
Quantile deviations (Figure~\ref{fig:quantile_agg}) show a positive deviation for the PTI Emulator at 3~h to 12~h aggregation at the least extreme quantile (0.8), shifting to increasingly strong underestimation at higher quantiles and longer aggregation windows up to three days, where the underestimation starts decreasing again. The AR Emulator instead shows a localized overestimation of up to +22.25\% at the 0.9999 quantile at 1~day aggregation, with underestimation confined to beyond 1~day in the most extreme quantile, but less strong than in the PTI case. Both modes show a good fit at the native 1 hour resolution.

\subsection{Temporal Evolution of Ensembles}

\begin{figure}[!htbp]
    \centering
    \includegraphics[width=1\linewidth]{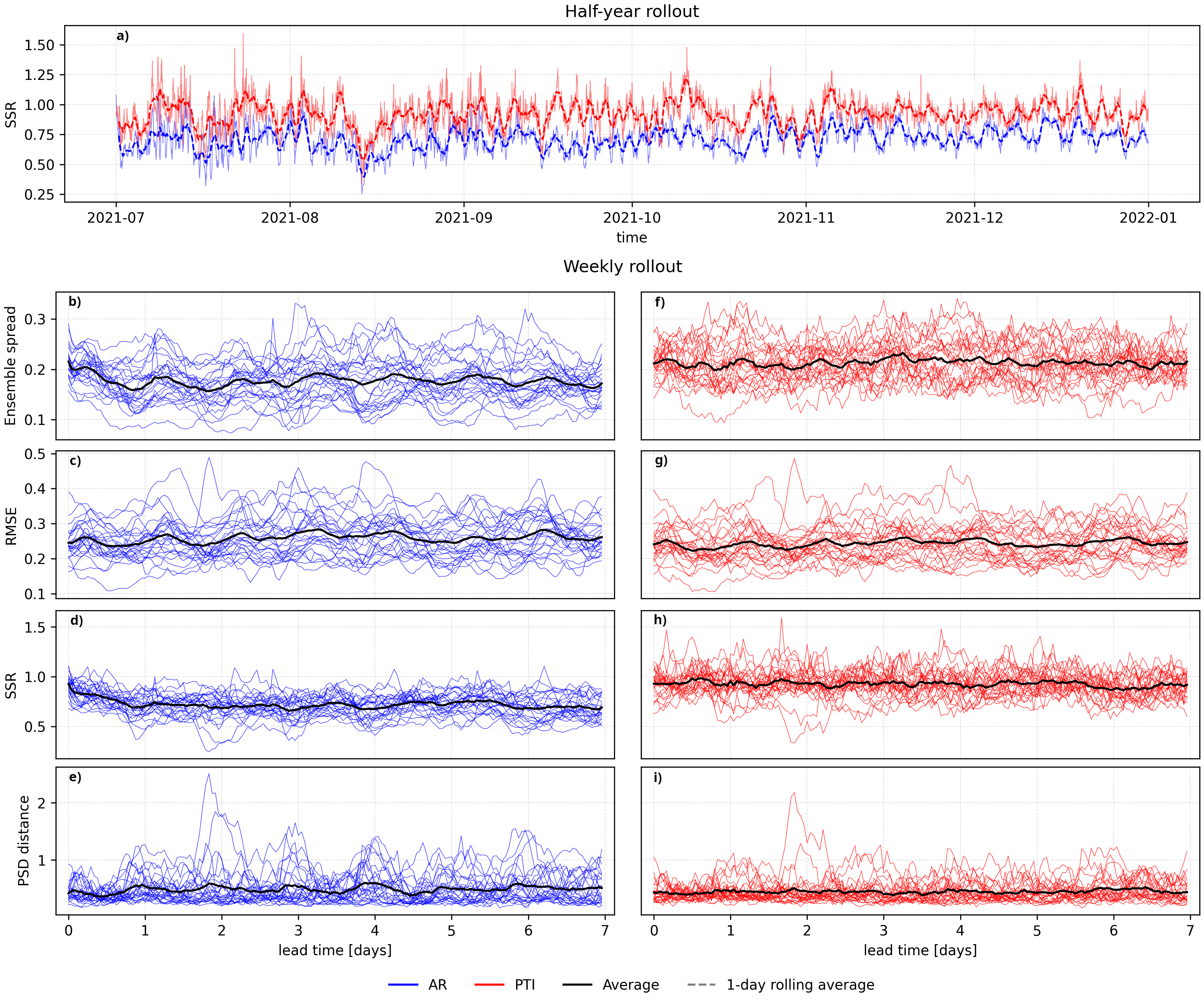}
    \caption{(a) Domain-averaged spread--skill ratio relative to RCM over a continuous, non-reinitialized half-year rollout (thin lines), together with 1-day rolling averages (dashed lines), for the AR and PTI Emulators.
(b–i) Diagnostics as a function of lead time for weekly reinitialized 7-day rollouts, showing individual initializations (thin lines) and their across-initialization mean (black lines). Ensemble spread (b, f), RMSE (c, g), Spread--skill ratio (d, h), and PSD distance averaged across ensemble members (e, i) are shown for the AR emulator (left) and PTI Emulator (right), respectively.
}
    \label{fig:rollout}
\end{figure}

\begin{figure}[!htbp]
    \centering
    \includegraphics[width=1\linewidth]{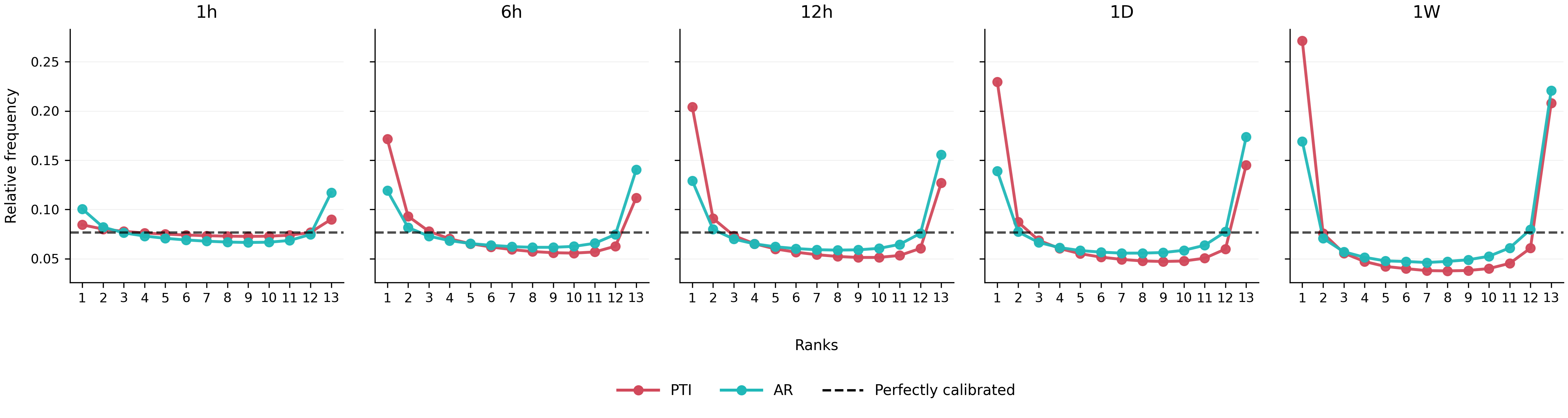}
    \caption{Rank histograms of precipitation comparing the 12-member ensembles generated by the AR and PTI Emulator modes against the RCM reference across different temporal aggregation scales.
}
    \label{fig:rank_hist}
\end{figure}

Figure~\ref{fig:rollout} shows uninterrupted (top panel) and weekly reinitialized (bottom panels) rollouts for the same 6 month test period for both modes. 
\\\\
The top panel shows the spread--skill ratio to RCM for both AR and PTI Emulators, each as a continuous time series and as 1-day rolling average. 
The rolling average spread--skill series of both modes are well correlated and fluctuate between 0.5 and 1.25 with the PTI Emulator sitting consistently higher. 
The lower panels show individual weekly evaluation segments (thin lines) of the test period. For the AR Emulator (left column) each trajectory is initialized at the beginning of a week (lead time 0) and evolves over the subsequent days. For the PTI Emulator (right column), predictions are generated independently at each time step, without temporal feedback between successive predictions. The added across-initialization average (black) for ensemble spread, RMSE, spread--skill ratio and PSD distance to RCM allow for a better direct comparison between modes. Both modes start from comparable spread at initialization by design, but the AR Emulator's spread declines over the first $\approx$24~h before stabilizing at a lower level, while the PTI Emulator's does not show this initial decline and overall higher averaged spread (AR Emulator: 0.18, PTI Emulator: 0.21). RMSE does not grow initially for either mode and is slightly higher throughout for the AR Emulator (0.256 vs.\ 0.244). Both modes reproduce the same diurnal pattern in RMSE. As a consequence of the spread difference, the PTI Emulator's spread--skill ratio sits closer to 1 (0.92 vs.\ 0.72).
\\\\
The PSD distance of the AR Emulator shows a slight increase during the first $\approx$24~h, followed by a stabilization at a relatively constant level. The PTI Emulator does not exhibit this initial increase. Overall, the average PSD distance is slightly higher for the AR Emulator (0.48) than for the PTI Emulator (0.44).
\\\\
To estimate ensemble calibration, Figure~\ref{fig:rank_hist} shows rank histograms (12 members) at five temporal aggregations (1~h, 6~h, 12~h, 1~day, 1~week) for both modes against the RCM. At 1~h , both histograms lie close to the uniform, perfectly calibrated reference, with the PTI mode showing a slightly closer alignment and a modest upward tilt at the highest rank. Underdispersion (frequency concentrated at the lowest and highest ranks) increases with aggregation for both modes, but more strongly for PTI. In addition, only PTI develops a pronounced tilt (a frequency decrease toward the higher ranks, not mirrored at the low end), which becomes visually distinct by 1~day and is most pronounced at 1~week.

\section{Discussion}

\subsection{Feasibility and Control: A Single Model Spanning AR and PTI Generation}

This subsection addresses the feasibility and control research question. A single trained model is shown to produce coherent AR-like rollouts (Feasibility) precisely because the single scalar noise level $\tau$ in the temporal-conditioning channel exposes that capability as a continuously selectable regime (Control) reproducing the full range of behavior between coherent autoregressive (AR) generation and independent per-timestep (PTI) generation without retraining. 
\\\\
The autocorrelation analysis (Figure~\ref{fig:acf_noise}) provides the clearest evidence for the level of control: varying $\tau$ traces a smooth, monotonic, non-crossing interpolation between the two regimes at every lag. Combined with the case study and spectral/extremes results above, this indicates that the noise-conditioned teacher-forcing scheme enables the model to learn how to modulate its reliance on the previous timestep according to its noise level, thereby providing a continuous means of controlling temporal dependence during inference.
\\\\
This design is a practical alternative to other strategies for obtaining temporally coherent output avoiding the joint prediction of larger space--time blocks and consecutive tiling while providing its own mechanism to initialize the model with its own PTI predictions. 
Our approach is keeping both the AR pathway and an equally viable PTI pathway available.
Inference remains per-timestep when requiring full parallelization, robustness to missing or irregular previous input timesteps, while AR feedback ensures strong temporal coherence. Parallel computation can still be achieved by adjustable temporal reset intervals or when generating multiple ensemble members independently. AR's own cost is architectural rather than purely statistical: conditioning on a previous generated state makes inference inherently sequential, forfeiting full parallelization and making the regime brittle whenever a previous state is missing or irregular --- precisely the scenario the PTI pathway is retained to cover. Retaining PTI prediction also matters because AR's benefits are not free as the next two subsections detail.
\\\\
Neither the diffusion backbone nor the noise-conditioned teacher-forcing scheme used to obtain this controllable AR--PTI spectrum is specific to our model architecture or to precipitation/temperature downscaling: both are generic components that can be substituted independently. This has implications for the interpretation and transferability of our approach.
\\\\
The specific characteristics and limitations identified in the following subsections reflect the particular model architecture, training setup, and data used in this study, rather than being inherent to the proposed approach of continuously controlling temporal dependence. They therefore remain open to improvement via architecture choice, loss design, or post-hoc calibration without forgoing the AR/PTI switch. At the same time, because the underlying mechanism only requires exposing a model during training to previous ground-truth states corrupted with randomly sampled noise levels, we see no obstacle to applying the same noise-conditioned teacher-forcing scheme to a broader range of sequential modeling tasks beyond the one studied here, and we would encourage such transfer.

\subsection{Autoregressive Rollout Robustness and Temporal Fidelity}
A central concern for any AR scheme is whether coherence is bought at the price of a characteristic AR failure mode: exposure bias compounding into error accumulation, progressive smoothing, or collapsing ensemble spread over long rollouts. PTI sidesteps this risk entirely by construction, having no closed feedback loop in which error could accumulate. 
The relevance of this concern is illustrated by the conventional teacher-forcing baseline (Section~\ref{sec:accumulation}), whose rollout diverges even though it starts from an observed RCM state. This contrasts with \citet{huang_generative_2026}, whose teacher-forced autoregressive emulator operates on daily data without such instability. We hypothesize that the temporal resolution explains this difference. At hourly resolution, consecutive states are strongly correlated, so that the uncorrupted previous state carries substantially more predictive information than the concurrent GCM state. Training then effectively reduces to a nowcasting problem, in which the network can minimize the loss largely by propagating the previous state and learns to rely only weakly on the GCM. The model therefore never learns to discount erroneous conditioning values, and once its own predictions are fed back, errors amplify without the GCM forcing pulling the rollout back. At daily resolution, the previous state is less informative relative to the GCM, which then remains the dominant predictor and anchors the rollout. This hypothesis remains to be tested, e.g.\ by training the baseline on daily data. The noise-conditioned teacher-forcing scheme counteracts this shortcut, since strongly corrupted previous states during training force the network to extract the relevant information from the GCM.
\\\\
Our half-year and weekly-reinitialized rollout analyses (Figure~\ref{fig:rollout}) find no substantial evidence of progressive error accumulation or degradation for the noise-conditioned AR mode (see animation of the half-year rollout in ancillary files). Over the half year rollout, the SSR trend is likely driven by seasonal change as it is observed for both modes. RMSE does not grow progressively over an initial 7-day lead time for either mode.
The AR mode's spread contracts once, over the first $\approx$24~h after initialization, before stabilizing at a constant level for the remainder of the rollout.
Similarly, the PSD distance shows a slight loss of fidelity over the first rollout hours in the AR mode compared to PTI, however, this is not visible in the case study or the 1-hour resolution histogram. Temporal autocorrelation provides a complementary indication of rollout stability, with AR closely reproducing the RCM autocorrelation structure and its decay towards negligible correlations at longer lags.
Small deviations nevertheless emerge at shorter timescales and add a further nuance: at low $\tau$, AR slightly underestimates autocorrelation at a 1-h lag but becomes slightly overly persistent at lags of 3--12~h  for both precipitation and temperature. This redistribution of temporal dependence may contribute to the changes in precipitation distributions under temporal aggregation. In particular, the slight overestimation of intermediate precipitation quantiles alongside well-represented or slightly underestimated upper quantiles at aggregation scales from 3~h  to 1 week is consistent with more persistent moderate precipitation rather than an amplification of precipitation extremes, without substantial loss of spatial fidelity as reflected by the relatively well-preserved RAPSD.
\\\\
For the noise-conditioned AR mode, exposure bias thus manifests as a regime-independent adjustment likely driven by the lower perturbation at a lower noise level $\tau$, not as the cumulative degradation seen in the TF baseline. The adjustment is confined to the first few steps and overall only leads to negligible degradation consistent with the findings of \citet{huang_generative_2026}, but, in contrast, does not necessitate discarding a spin-up period, as the main notable change is a contraction of the ensemble spread.
AR exhibits a lower SSR than PTI (0.72 vs. 0.92), this difference is also primarily driven by the spread offset rather than differences in RMSE. It therefore reflects underdispersion in AR at its native 1-hour resolution rather than a deficit in predictive skill.
\\\\
Critically, the modest 1-hour underdispersion of AR compared to a better calibrated PTI is more than offset once aggregation is considered: rank histograms show PTI's underdispersion worsening substantially under aggregation, exceeding AR's underdispersion. Ensemble spread at the model's native resolution represents a tunable model bias that may be further encouraged by training-time or post-hoc optimization, independent of the noise-conditioned teacher-forcing scheme. Correcting biases that emerge under temporal aggregation, however, is more challenging, precisely where AR's advantage over PTI becomes apparent.

\subsection{Downstream relevance}

Whether temporal consistency matters in practice hinges on its effect on quantities accumulated over time, since most downstream impacts --- fluvial flooding, soil saturation, heat-spell duration --- are governed by accumulation or persistence rather than instantaneous or independent, per-timestep values. As a proxy for downstream applicability, we thus analyzed the rainfall statistics of different temporal aggregations.
\\\\
A first indication of differences can be seen in the precipitation case-study: at the per-timestep (1~h) level, AR and PTI are visually comparable. The distinction emerges once fields are aggregated or considered in a sequential order. While AR preserves the temporal relationships between successive fields, allowing spatial patterns and precipitation features to evolve coherently, PTI's independently generated fields fail to capture these dependencies resulting in differences in the aggregated magnitude and spatial distribution. RAPSD results support this interpretation: the spectral gap between modes is indistinguishable at 1~h and opens only under aggregation, with PTI producing increasingly overly smooth aggregated fields.
\\\\
PTI underestimates the most extreme aggregated precipitation rates by up to a factor of two at 3~h to 3 days, while overestimating weaker precipitation rates and its otherwise well-calibrated ensemble also becomes increasingly underdispersive under temporal aggregation, with a more pronounced loss of calibration than in AR.
\\\\
These deficiencies are relevant to hydrological modeling, where runoff generation depends on the temporal and spatial evolution of rainfall events. More broadly, physical process models rely on dynamically consistent sequences of atmospheric states to represent interactions between processes and their cumulative effects. The absence of such consistency may therefore introduce errors that extend beyond biases in individual forcing variables shown here, potentially propagating through coupled processes and affecting the simulated system evolution.
\\\\
For temperature, the evidence for downstream relevance is considerably weaker at the given spatial and temporal scales. Differences in temperature-anomaly autocorrelation between modes are small, and in the case study, while PTI shows more unphysical, non-smooth anomaly changes between consecutive timesteps, both modes converge to indistinguishable behavior already by 12-hour accumulation, far sooner than the multi-day timescale at which precipitation modes reconverge. 
Both modes also faithfully reproduce the diurnal cycle, strongly constrained by the GCM input itself, a constraint far weaker for precipitation. 
\\\\
The temporal independence of PTI should affect the representation of extremes and derived quantities such as daily temperature maxima. There, a lack of persistence could increase the likelihood of isolated extreme values occurring within a given day. Here, we observe temperature-anomaly trajectories in the PTI mode that remain slightly implausible at the per-timestep level, but cannot robustly demonstrate a concrete downstream consequence within this study's scope.

\section{Conclusions}

This study explores feasibility, risks and benefits of autoregressive generative emulation of regional climate models applied in a GCM downscaling setting using a diffusion U-Net backbone and proposes a solution in the form of a noise-conditioned teacher-forcing scheme that allows for a choice between autoregressive and independent sampling at inference by a single trained model.
\\\\
Applying our method to a European-domain, hourly ICON-downscaled (RCM) historical storyline simulation of surface precipitation and 2-meter temperature in the AWI-CM1 GCM, we found that an adjustable noise parameter applied to the previous high resolution timestep provides a continuous transition between two sampling modes. At one end, autoregressive sampling closely follows RCM's autocorrelation function, while at the other, purely per time step independent generation is achieved with respect to the previous high-resolution state.
An investigation of the main risks of autoregressive diffusion models, namely smoothing, error accumulation and loss of diversity, caused by potential exposure-bias, found no severe and, most importantly no progressive degradation in the autoregressively generated predictions over increasing rollout lengths. The only notable observable change is a loss of ensemble spread over the first 24 hours that stabilizes at an acceptable level, not making a later-discarded spin-up period necessary.
A conventional teacher-forcing baseline trained on uncorrupted previous states, in contrast, diverged during the six-month rollout despite being initialized with an observed RCM state, which highlights the stabilizing role of the noise conditioning.
Nevertheless, the projected relevance for potential downstream application of the produced fields is real. 
While only negligible differences in hourly and daily accumulated temperature fields were found between autoregressive and independent generation, we observed an up to twofold underestimation of precipitation extremes and up to twofold overestimation of light rainfall in accumulated independently generated fields. The autoregressive mode tracks RCM's extreme event characteristics more closely throughout all accumulations showing only slight overestimation for medium rainfall and a significantly reduced underestimation of the highest quantile of the distribution.
\\\\
More broadly, our results demonstrate an elegant, architecture-agnostic solution to the problem of temporal inconsistency in diffusion-based downscaling, enabling autoregressive inference without substantial degradation over extended rollouts.
The proposed approach provides an adjustable transition between autoregressive rollouts and per-timestep independent sampling, allowing users to select the sampling strategy according to the requirements of their application.
The two sampling modes offer complementary advantages: autoregressive sampling ensures temporal consistency and better preserves accumulated precipitation extremes, while independent sampling provides a flexible means of initializing autoregressive rollouts without requiring an initial high-resolution target state. Additionally, independent sampling retains per-timestep parallelization and exhibits slightly better ensemble characteristics at native hourly resolution.
\\\\
Together with alternative approaches \citep{hess_generating_2025, huang_generative_2026} for temporal consistency, our solution for autoregressive inference provides good evidence that the field of data driven RCM emulation and data driven downscaling more generally can move on to not treating temporal consistency as an often out-of-scope added complexity, but an easily achievable necessity, for future model design.

\section*{Author Contribution Statement}
LG conceived and designed the study with help from JP and CC. The method development, coding, model training and evaluation were carried out by LG. Formal data analysis was performed by LG and JP. ASB and PL provided the global and regional simulation datasets. The first draft of the manuscript was written by JP and LG. All authors have reviewed and edited the manuscript. Funding acquisition and overall project supervision were managed by HK, BF, PL and CC. All authors have read and agreed to the published version of the manuscript.

\section*{Acknowledgement}
This work was supported by the Helmholtz Research Field Earth \& Environment for the Innovation Pool Project ACTUATE and the Helmholtz Foundation Model Initiative (HFMI) project HClimRep. The computational resources used for this study were provided by the Deutsches Klimarechenzentrum (DKRZ), granted by its Scientific Steering Committee (WLA) under project IDs 1264 (GCM and RCM data generation) and 1343 (deep learning experiments).

\section*{Code and data availability}
The data and code supporting this study will be made publicly available upon publication.

\section*{Use of AI}
Large language models were used to help with analysis code and refining the language. They were not used to introduce new intellectual content. The authors take full accountability for the analysis, figures, and text.

\bibliographystyle{plainnat}
\bibliography{references}

\end{document}